\documentclass[10pt,twocolumn]{IEEEtran}
\usepackage{amsmath}
\usepackage{algpseudocode,algorithm,algorithmicx}
\usepackage{hyperref}
\usepackage{graphicx}
\usepackage{amssymb}
\usepackage{amsmath}
\usepackage{cite}
\usepackage{epsfig}
\usepackage{stfloats}
\usepackage{subfigure}
\usepackage{psfrag}
\usepackage[mathscr]{euscript}
\usepackage{acronym}
\usepackage{booktabs}
\usepackage{epstopdf}
\usepackage{pstool}
\usepackage{array}
\usepackage{amsmath}
\usepackage{graphicx}
\DeclareGraphicsExtensions{.pdf,.png,.jpg}

\title{Orthogonal Code-based Block Transmission\\for Burst Transmission \vspace{0.5cm}}  

\author{
Hyejin~Kim$^\dagger$, Insik~Jung, \textit{Student Member,~IEEE}, Wonsuk~Chung, Sooyong~Choi, \textit{Member,~IEEE}, and Daesik~Hong$^\ddagger$, \textit{Senior Member,~IEEE}\vspace{-0.5cm}}

\begin{document}
\maketitle

\begin{abstract}
This paper proposes a new multi-carrier system, called orthogonal code-based block transmission (OCBT). OCBT applies a time-spreading method with an orthogonal code to have a block signal structure and a windowing procedure to reduce the out-of-band (OOB) radiation. The proposed OCBT can transmit the quadrature amplitude modulation (QAM) signals to use the conventional multiple input multiple output techniques. Numerical results show that the proposed OCBT using QAM signal has the short burst compared to the filter-bank multi-carrier (FBMC), the low complexity compared to FBMC and windowed orthogonal frequency division multiplexing (W-OFDM) and also the low OOB radiation compared to OFDM.
\end{abstract}

\section{Introduction} \label{Sec_Intro}
 A large number of multi-carrier schemes are available for broadband wireless communication systems. Among them, orthogonal frequency division multiplexing (OFDM) has been selected as a wireless transmission technology in 4G communication \cite{E5}. However, OFDM requires a cyclic prefix (CP) in multi-path channel and a sufficient guard band in the frequency domain due to the high out-of-band (OOB) radiation \cite{C3}.

 To enhance the spectral efficiency of the conventional CP-OFDM which requires the CP and the sufficient guard band, M.Bellanger proposed filter bank multi-carrier (FBMC) \cite{E2}. In order to reduce the OOB radiation, FBMC filters each sub-carrier signal individually with prototype filters which are well-localized in the frequency and time domains.  Therefore FBMC has the reduced guard band and does not need the CP.

 However, FBMC still has the long burst due to the transition time caused by the overlap and sum structure. In addition, FBMC has to use offset quadrature amplitude modulation (OQAM) to avoid the intrinsic interference \cite{J4}. Therefore, it is difficult for FBMC to use the conventional QAM based multiple input multiple output (MIMO) techniques \cite{J9}.

 As an approach to solving the problems that FBMC has the long burst and difficulties in applying the conventional QAM based MIMO techniques, we propose a new multi-carrier system, orthogonal code-based block transmission (OCBT). In OCBT, we use a time-spreading method and divide the spread symbols in the orthogonal code domain. Then, we can overlap the spread symbols without being staggered, making the signal a block structure and transmitting a QAM signal. Also, we apply the windowing procedure to reduce the OOB radiation.

\section{OCBT Transceiver Design} \label{Sec_Transceiver}

   \begin{figure*}[b!]
   \center{
   \subfigure[OCBT transceiver]
   {     \includegraphics[width=1\columnwidth]{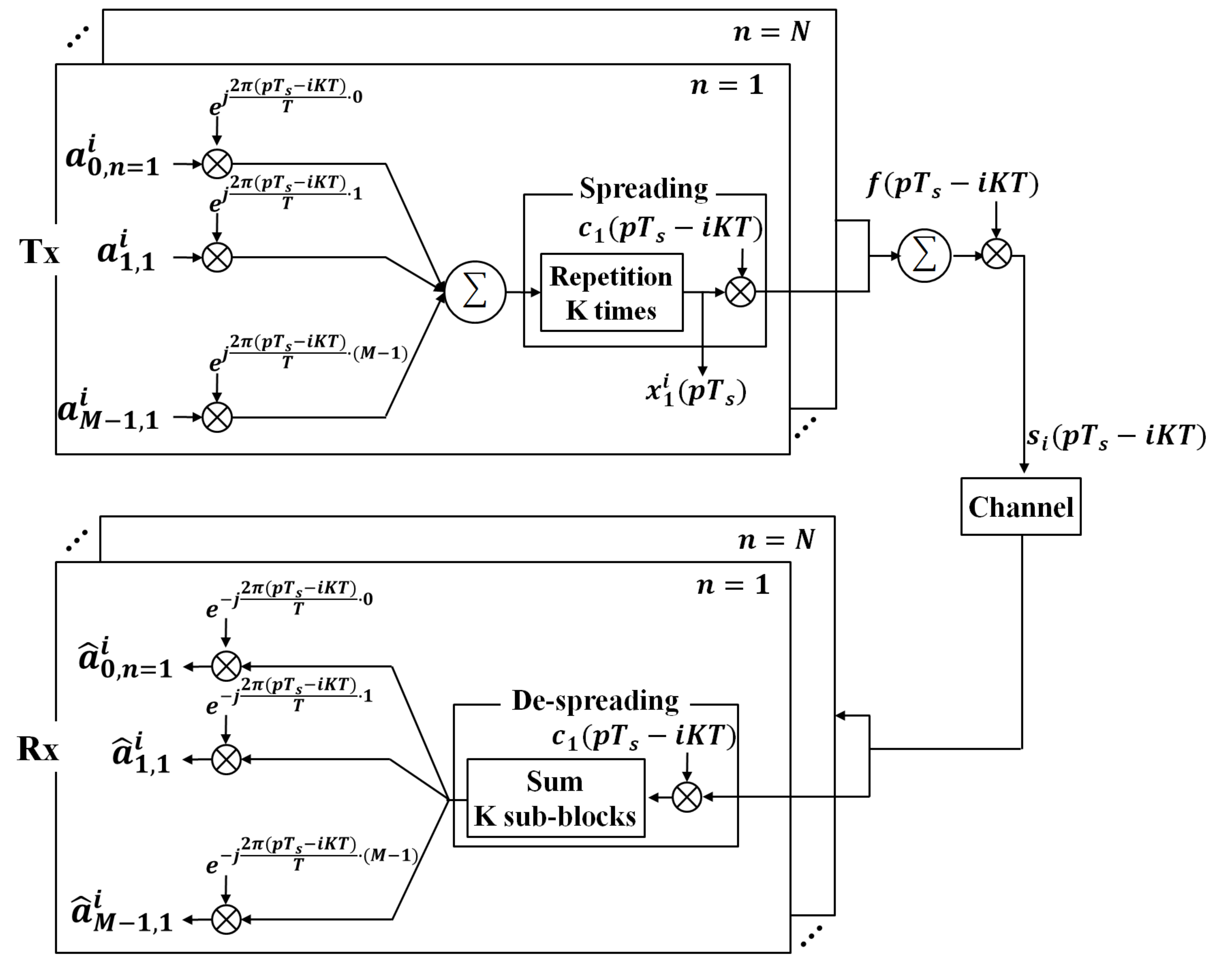}
         \label{Fig_OCBT_transceiver}
    }
    \vspace{-0.1cm}
   \,
    \subfigure[PPN-FBMC transceiver]
   {
        \includegraphics[width=0.92\columnwidth]{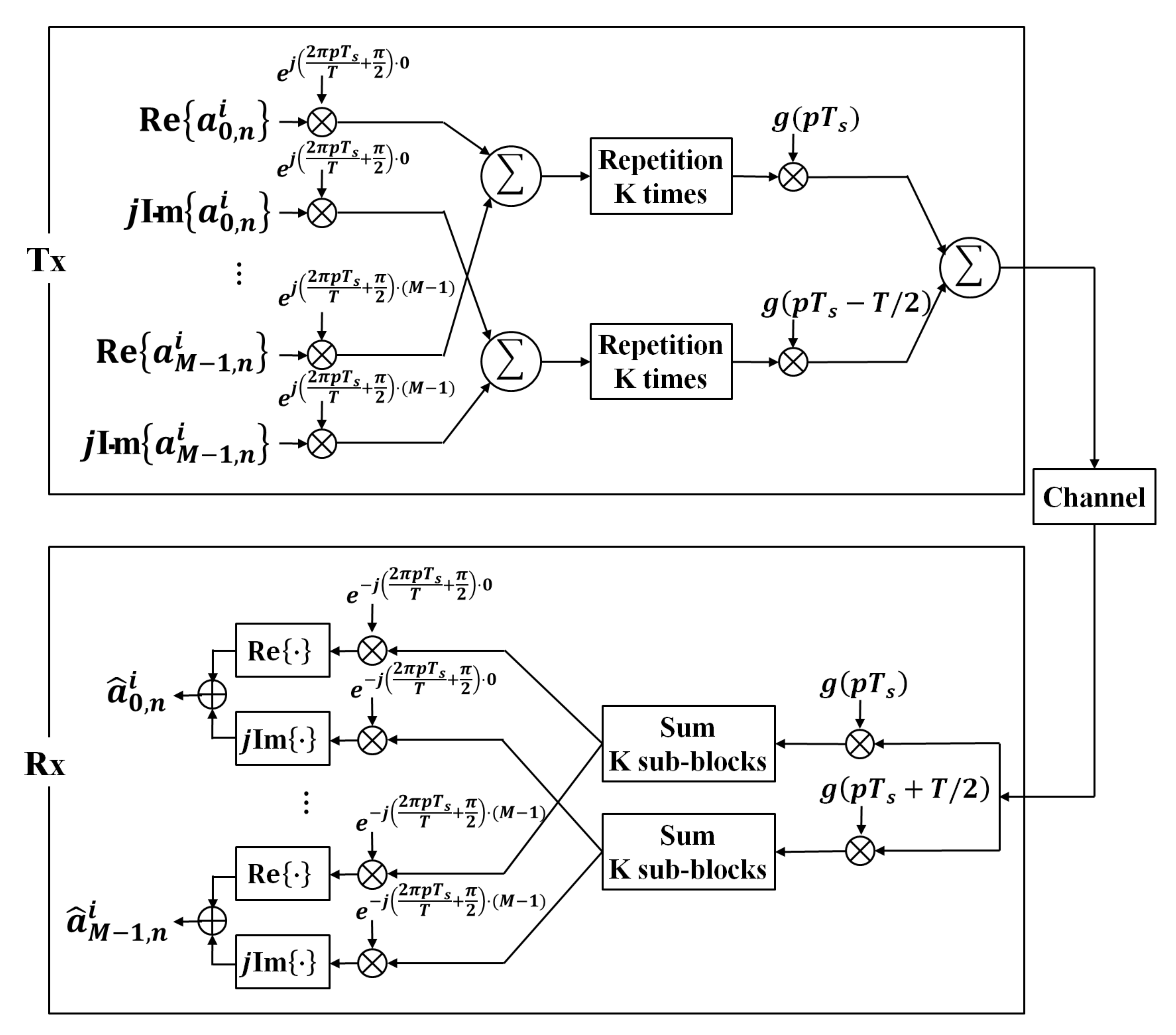}
        \label{Fig_FBMC_transceiver}
        }
        \vspace{-1mm}
   \caption{OCBT and FBMC transceivers.}
   \vspace{-0.1cm}
   \label{Fig_Transceiver}
   }
   \end{figure*}

\subsection{Transmitter} \label{Subsec_Transmitter}

   In this section, we will introduce the proposed OCBT shown in Fig. \ref{Fig_Transceiver}\subref{Fig_OCBT_transceiver} by comparing it with the conventional poly-phase network (PPN)-FBMC shown in Fig. \ref{Fig_Transceiver}\subref{Fig_FBMC_transceiver} \cite{C5}.

    The block structure of OCBT consists of $N$ symbols ($N \le K$, $K$ is the spreading factor) and each symbol has $M$ sub-carriers. The transmitted OCBT signal $s(p{T_s})$ can be expressed as \cite{J4}
    \begin{equation}
    \label{Eq_transmit_signal}
    s(p{T_s})=\sum\limits_{i =  - \infty }^\infty{s_i(p{T_s}-iKT)},
    \end{equation}
    where $p$ and $T_s$ are the sample index and the sample time, respectively. $s_i(p{T_s})$ is the transmitted block signal in the $i$-th signaling time of length $KT$, where $T=M{T_s}$.

    Here, we consider the generation of the $i$-th block signal $s_i(p{T_s})$, as shown in Fig. \ref{Fig_Transceiver}\subref{Fig_OCBT_transceiver}. Let us denote the $n$-th QAM data on the $m$-th sub-carrier in the $i$-th signaling time as $a_{m,n}^i$ ($n=1,2,...,N$, $m=1,2,...,M$) which is the element of the data vector ${\bf{a}}_n$. As depicted in Fig. \ref{Fig_Transceiver}\subref{Fig_OCBT_transceiver}, $a_{m,n}^i$ in OCBT passes through the inverse FFT (IFFT) without separation of the real and the imaginary parts, compared to FBMC where the real and imaginary parts of $a_{m,n}^i$ pass through the IFFT separately, as shown in Fig. \ref{Fig_Transceiver}\subref{Fig_FBMC_transceiver}. After the IFFT, we use the time-spreading method by orthogonal code. For the time-spreading, we begin by repeating the IFFT outputs $K$ times, similarly to the approach with PPN-FBMC. Let us call the $M$-point IFFT output a sub-block. The result of repeating sub-block $K$ times is defined as a symbol $x_n^i(p{T_s})$, which is element of the vector ${\bf{x}}_n$. The $n$-th symbol vector ${\bf{x}}_n$, which consists of the $K$ sub-blocks, is then written as
   \begin{equation}
   \label{Eq_xcf}
   \begin{array}{l}
   {\bf{x}}_n \buildrel \Delta \over = [x_{n,0}, x_{n,1}, ..., x_{n,KM-1}]^T\\
    \quad\,\;= {\bf{R}}{{\bf{W}}_M^H}{{\bf{a}}_n},
    \end{array}
   \end{equation}
   where $x_{n,p}$ is the $p$-th sample in the symbol $x_n^i(p{T_s})$. ${{\bf{W}}_M^H}$ is the $M$$\times$$M$ inverse FFT (IFFT) matrix which is the Hermitian of the FFT matrix and ${\bf{R}}$ is the $KM$$\times$$M$ repetition matrix defined as
   \begin{equation}
   \label{Eq_duplication}
   {\bf{R}} = { {{{\left[ {{\bf{I}}_M}, {{\bf{I}}_M}, \cdots {{\bf{I}}_M} \right]}^T}}},
   \end{equation}
   where ${{\bf{I}}_M}$ is the $M \times M$ identity matrix and ${\{ \cdot \}^T}$ is the transpose of $\{ \cdot \}$.

   Then, we overlap the $N$ symbols $x_n^i(p{T_s})$ of length $KT$ to prevent the spectral efficiency from being degraded by repetition. Unlike FBMC which has the transition time caused by the overlap and sum structure using the orthogonal characteristic of prototype filter $g(p{T_s})$ \cite{E2}, OCBT uses the orthogonal codes ${c_n}(p{T_s})$, represented by the $KM \times KM$ diagonal matrix ${\bf{C}}_n$, to overlap the symbols completely without transition time. After the symbols $x_n^i(p{T_s})$ are multiplied by the orthogonal codes, $N$ spread signals ${c_n}(p{T_s}-iKT){x_n^i}(p{T_s})$ are added to make a block structure with the normalization.

   The $N$ orthogonal codes ${c_n}(p{T_s})$ are made with the corresponding $K$-point orthogonal code sequences ${c_{n,k}}$ ($k=0,1,...,K-1$) which satisfy the orthogonal condition as
   \begin{equation}
   \label{Eq_code_orthogonality}
   \sum\limits_{k = 0}^{K-1} {c_{n,k}c_{l,k}}  = \left\{ {\begin{array}{*{20}{c}}
    K\\
    0
    \end{array}{\rm{      }}} \right.\begin{array}{*{20}{c}}
    {n = l}\\
    {n \ne l}
    \end{array}.
   \end{equation}
   We define non-zero elements of ${c_n}(p{T_s})= c_{n,\left\lfloor p/M \right\rfloor}$ in $[0,KT]$, where $\left\lfloor \cdot \right\rfloor$ is a floor function. In other words, multiplying $x_n^i(p{T_s})$ by ${c_n}(p{T_s}-iKT)$ means that the $k$-th sub-blocks of ${x_n^i}(p{T_s})$ are multiplied by ${c_{n,k}}$. In this paper, we use the Walsh code as an example of the orthogonal code.

   The signal is then transmitted after being multiplied by the window $f(p{T_s})$, represented by $KM \times KM$ diagonal matrix ${\bf{F}}$, to reduce the side-lobe level of the frequency response such as Windowed-OFDM (W-OFDM) \cite{C4}. In $f(p{T_s})$ where non-zero elements are defined in $[0,KT]$, a window $f_p$ of length $T$ is repeated $K$ times to maintain the repetition structure of the spread signal and satisfy the code orthogonality. Finally, the transmitted signal in the $i$-th signaling time ${\bf{s}}_i$ can be written as \cite{C2}
   \begin{equation}
   \label{Eq_transmit_signal_mat}
   \begin{array}{l}
   {\bf{s}}_i \buildrel \Delta \over = \left[ s_{i,0}, s_{i,1}, ...,s_{i,KM-1}\right]^T\\
   \quad\, = \frac{1}{{\sqrt N }}{\bf{F}}\sum\limits_{n = 1}^N {{{\bf{C}}_n}{\bf{x}}_n}\\
   \quad\, = \frac{1}{{\sqrt N }}{\bf{F}}\sum\limits_{n = 1}^N {{{\bf{C}}_n}{\bf{R}}{{\bf{W}}_M^H}{{\bf{a}}_n}},
   \end{array}
   \end{equation}
   where $s_{i,p}$ denotes the $p$-th sample in the $i$-th block signal $s_i(p{T_s})$.

\subsection{Receiver} \label{Subsec_Receiver}

   The receiver for OCBT is depicted in Fig. \ref{Fig_Transceiver}\subref{Fig_OCBT_transceiver}. The received signal $y(p{T_s})$ consists of the convolution of the impulse response of the channel $h(p{T_s})$ and the transmitted signal $s(p{T_s})$ in (\ref{Eq_transmit_signal}) and the noise $\eta (p{T_s})$ as
   \begin{equation}
   \label{Eq_received_signal}
   y(p{T_s}) = h(p{T_s}) * s(p{T_s}) + \eta (p{T_s}).
   \end{equation}
   Let us denote the received signal $y(p{T_s})$ in the $i$-th signaling time as $y_i(p{T_s})$. $y_i(p{T_s})$ consists of the $i$-th block transmission signal and the ($i-1$) block transmission signal because of the channel effect. A vector ${\bf{y}}_i$ whose elements are $KM$ samples of $y_i(p{T_s})$ can then be obtained as \cite{J7}
   \begin{equation}
   \label{Eq_received_signal_mat}
   {\bf{y}}_i={{\bf{h}}_i}{{\bf{s}}_i}+{{\bf{h}}_{i-1}}{{\bf{s}}_{i-1}}+ {\boldsymbol{\eta}},
   \end{equation}
   where ${\boldsymbol{\eta}}$ is the noise vector. ${\bf{h}}_i$ denotes the $KM$$\times$$KM$ lower triangular Toeplitz matrix with $[h(0\cdot{T_s}),...,h((G-1)\cdot{T_s}),0,...,0]^T$ in the first column, and ${\bf{h}}_{i-1}$ denotes the $KM$$\times$$KM$ upper triangular Toeplitz matrix with $[0,...,0,h((G-1)\cdot{T_s}),...,h(1\cdot{T_s})]$ in the first row where the channel coefficient $h(p{T_s})$ has the $G$ taps.

   Firstly, the received signal is equalized in the frequency domain to cancel the channel effect. After channel equalization, the received signal in the time domain is expressed as
   \begin{equation}
   \label{Eq_channel_equalization}
   {{\bf{\tilde y}}_i}={{\bf{W}}_{KM}^H}{\bf{H}}_{EQ}{{\bf{W}}_{KM}}{{\bf{y}}_i}\\
   ={{\bf{s}}_i}+{{\bf{\tilde h}}_{i-1}}{{\bf{s}}_{i-1}}+ {\boldsymbol{{\tilde \eta}}},
   \end{equation}
   where ${\bf{W}}_{KM}$ and ${\bf{W}}_{KM}^H$ are the $KM \times KM$ FFT and IFFT matrix, respectively. ${\bf{H}}_{EQ}$ is a $KM$$\times$$KM$ equalization matrix in \cite{J10} using the frequency response of the channel such as zero forcing and minimum mean square error (MMSE) equalizers. ${{\bf{\tilde h}}_{i-1}}$ and ${\boldsymbol{{\tilde \eta}}}$ are the modified ${{\bf{h}}_{i-1}}$ and ${\boldsymbol{\eta}}$ after the channel equalization, respectively.

   Then, to de-spread the data ${\bf{a}}_l$ among the $N$ symbols, the received signal after channel equalization ${{\bf{\tilde y}}_i}$ is multiplied by the corresponding code ${\bf{C}}_l$. The data ${\bf{a}}_l$ can be recovered by summing the $K$ spread sub-blocks and performing the $M$-point FFT reversing the transmission process in (\ref{Eq_transmit_signal_mat}). In de-spreading, $\frac{\sqrt N}{K}$ has to be multiplied for normalization. From (\ref{Eq_transmit_signal_mat}) and (\ref{Eq_channel_equalization}), the recovered data $\widehat {{{\bf{a}}_l}}$ for ${\bf{a}}_l$ can be obtained as follows:
   \begin{equation}
   \label{Eq_recovered_data}
   \begin{array}{l}
    \widehat{{\bf{a}}_l} = {{\bf{W}}_M}\frac{{\sqrt N }}{K}{{\bf{R}}^T}{{\bf{C}}_l}{{{\bf{\tilde y}}}_i}\\
   \quad\,= {{\bf{W}}_M}\frac{{\sqrt N }}{K}{{\bf{R}}^T}{{\bf{C}}_l}\left( {{{\bf{s}}_i} + {{{\bf{\tilde h}}}_{i - 1}}{{\bf{s}}_{i - 1}} + {\boldsymbol{{\tilde \eta}}} } \right)\\
   \quad\, = {{\bf{W}}_M}\frac{{\sqrt N }}{K}{{\bf{R}}^T}{{\bf{C}}_l}{{\bf{s}}_i} + IBI + {\bf{N}}\\
   \quad\, = {{\bf{W}}_M}\frac{{\sqrt N }}{K}{{\bf{R}}^T}{{\bf{C}}_l}\left(\frac{1}{{\sqrt N }}{\bf{F}}\sum\limits_{n = 1}^N {{{\bf{C}}_n}{\bf{R}}{{\bf{W}}_M^H}{{\bf{a}}_n}}\right)+IBI+{\bf{N}},
      \end{array}
   \end{equation}
   where ${{\bf{R}}^T}$ is a $M$$\times$$KM$ matrix which sums the $K$ spread sub-blocks. Inter-block interference (IBI) is the interference received in one symbol from the previous block, and ${\bf{N}}$ denotes the modified noise vector after de-spreading.

    \section{Performance Analysis}\label{Sec_Performance_Analysis}

    In this section, we analyze the performances of OCBT structure with the matrix model.

    \subsection{ISI-free Structure}\label{Subsec_Orthogonal_Condition}

   As mentioned in Section \ref{Subsec_Transmitter}, the window $f_p$ of length $T$ is $K$ times repeated in the window $f(p{T_s})$ to maintain the repeated structure of the spread signal and satisfy the code orthogonality. That is, the window matrix ${\bf{F}}$ in (\ref{Eq_recovered_data}), whose diagonal elements are $f(p{T_s})$, also has $K$ times repeated diagonal elements $f_p$, so ${\bf{F}}$ does not destroy the orthogonality between the code matrices.
   Thus, ${{\bf{R}}^T}{{\bf{C}}_l}{\bf{F}}{{\bf{C}}_n}{{\bf{R}}}$ in (\ref{Eq_recovered_data}) equals $\left( {\sum\limits_{k = 0}^{K - 1} {{c_{l,k}}{c_{n,k}}} } \right){{\bf{F}}_M}$ which is ${\bf{0}}$ if $n \ne l$ from (\ref{Eq_code_orthogonality}), where ${{\bf{F}}_M}$ is a diagonal matrix whose diagonal elements are $M$ samples of $f_p$. As a result, the other ($N-1$) data ${{\bf{a}}_n}$ ($n=1,2,...,N, n \ne l$) are cancelled, and the recovered data in (\ref{Eq_recovered_data}) can be written as
   \begin{equation}
   \label{Eq_ISI_free}
    \widehat{{\bf{a}}_l} = {{\bf{W}}_M}{{\bf{F}}_M}{{\bf{W}}_M^H}{{\bf{a}}_l}+IBI+{\bf{N}}.
   \end{equation}
   In other words, the orthogonal code in OCBT makes no inter-symbol interference (ISI) from the overlapped symbols in the same block.

   \subsection{IBI Reduction}\label{Subsec_IBI_Reduction}

   After channel equalization, the interference from the previous block in (\ref{Eq_channel_equalization}) is expressed as
   \begin{equation}
   \label{Eq_IBI_aft_ch_eq}
    {{{\bf{\tilde h}}}_{i - 1}}{{\bf{s}}_{i - 1}} = {\left( {{{\left[ {\begin{array}{*{20}{c}}
    {{\boldsymbol{\epsilon}^T _{i - 1,1}}}& {{\boldsymbol{\epsilon}^T _{i - 1,2}}} & \cdots &{{\boldsymbol{\epsilon}^T _{i - 1,K}}}\end{array}} \right]}^T}} \right)_{KM \times 1}},
    \end{equation}
   where $\boldsymbol{\epsilon} _{i - 1,k}$ having $M$ elements is the $k$-th sub-block vector of the interference from the ($i-1$)-th block after channel equalization.
   From triangle inequality, the interference power after de-spreading in (\ref{Eq_recovered_data}) is reduced as
   \begin{equation}
   \label{interference_P}
    \begin{array}{l}
    {\left\| {\frac{{\sqrt N }}{K}{{\bf{R}}^T}{{\bf{C}}_l}{{{\bf{\tilde h}}}_{i - 1}}{{\bf{s}}_{i - 1}}} \right\|^2}\\
    \quad\qquad = {\left\| {\frac{{\sqrt N }}{K}\left( {{{{\boldsymbol{ \epsilon }}}_{i - 1,1}} \pm {{{\boldsymbol{ \epsilon }}}_{i - 1,2}} \pm  \cdots  \pm {{{\boldsymbol{ \epsilon }}}_{i - 1,K}}} \right)} \right\|^2}\\
    \quad\qquad\le {\left( {\frac{{\sqrt N }}{K}\left( {\left\| {{{{\boldsymbol{ \epsilon }}}_{i - 1,1}}} \right\| + \left\| {{{{\boldsymbol{ \epsilon }}}_{i - 1,2}}} \right\| \cdots  + \left\| {{{{\boldsymbol{ \epsilon }}}_{i - 1,K}}} \right\|} \right)} \right)^2}\\
    \quad\qquad\le \frac{N}{{{K^2}}}\left( {{{\left\| {{{{\boldsymbol{ \epsilon }}}_{i - 1,1}}} \right\|}^2} + {{\left\| {{{{\boldsymbol{ \epsilon }}}_{i - 1,2}}} \right\|}^2} \cdots  + {{\left\| {{{{\boldsymbol{ \epsilon }}}_{i - 1,K}}} \right\|}^2}} \right)\\
    \quad\qquad= \frac{N}{K^2}{\left\| {{{{\bf{\tilde h}}}_{i - 1}}{{\bf{s}}_{i - 1}}} \right\|^2}.
    \end{array}
   \end{equation}
   If the time-spreading method is not applied ($K=1, N=1$), the interference is spread in one symbol after de-spreading in the receiver. In OCBT, however, the interference is spread to $N$ symbols in $K$ sub-blocks after de-spreading. Thus, from (\ref{interference_P}), the power of the interference received by one symbol, that is IBI, is reduced by ${N}/{K^2}$ times or more.
   Moreover, as $K$ increases, the interference is spread to more sub-blocks, thus reducing the IBI even more. Therefore, OCBT can be robust for multi-path channels even without adding overhead.

    \subsection{SINR Analysis}\label{Subsec_SINR_Analysis}

   ${{\bf{W}}_M}{{\bf{F}}_M}{\bf{W}}_M^H$ in (\ref{Eq_ISI_free}) is not equal to ${{{\bf{I}}_M}}$, and the $(u,v)$-th element of it is expressed as
   \begin{equation}
   \label{Eq_WFW}
   {\left( {{{\bf{W}}_M}{{\bf{F}}_M}{\bf{W}}_M^H} \right)_{uv}} = {\left(\frac{1}{M}\sum\limits_{p = 0}^{M - 1} {{f_{p + 1}}{\omega ^{(u - v)p}}}\right)},
   \end{equation}
   where $\omega={e^{- j\frac{{2\pi}}{M}}}$. Thus, windowing makes the inter-carrier interference (ICI). From (\ref{Eq_ISI_free}) and (\ref{Eq_WFW}), we obtain the received data $\widehat {{{\bf{a}}_l}}$ on the $m$-th sub-carrier as follows:
   \begin{equation}
   \label{Eq_received_symbol_2}
   \begin{array}{l}
    {{\widehat a}_{m,l}}\\
    = \underbrace {\frac{1}{M}\sum\limits_{\scriptstyle p = 0 \atop {\scriptstyle \atop \scriptstyle}}^{M - 1} {{f_{p + 1}}} {a_{m,l}}}_{\scriptstyle {E\left[ {{f_p}} \right]{a_{m,l}}}} + \underbrace {\frac{1}{M}\sum\limits_{\scriptstyle m' = 1 \atop \scriptstyle m' \ne m}^M {\sum\limits_{p = 0}^{M - 1} {{f_{p + 1}}{\omega ^{(m - m')p}}} {a_{m',l}}} }_{\scriptstyle {ICI_{m,l}}}\\[30pt]
    \quad+ {\textstyle {IBI_{m,l}}} + {{N}}_{m,l},
    \end{array}
   \end{equation}
   where ${{a}_{m,l}}$ and ${{a}_{m',l}}$ are the data ${{\bf{a}}_l}$ on the $m$-th and $m'$-th sub-carrier, and ${ICI}_{m,l}$, ${IBI}_{m,l}$ and ${{N}}_{m,l}$ are the ICI, IBI and noise of the $l$-th data on the $m$-th sub-carrier, respectively.
   From (\ref{Eq_received_symbol_2}), the data in each sub-carrier is received as a sum of the desired data which decreases to the average value of the window $f_p$, ICI from the other $M-1$ sub-carriers, IBI from the previous block and noise.
   Consequently, the signal to interference plus noise ratio (SINR) is expressed as
   \begin{equation}
   \label{eq_SINR}
   SINR = \frac{{{E^2}\left[ {f_p} \right]P_s}}{{ICI + IBI + P_N}},
   \end{equation}
   where $P_s$ and $P_N$ denote the power of the signal and the noise, respectively.

    The window can reduce the OOB radiation \cite{C4}, however, the received symbol can be distorted due to the ICI caused by windowing as (\ref{Eq_received_symbol_2}). In \cite{J8}, the window even experiences a trade-off between frequency confinement and signal distortion caused by ICI. In other words, the window can be a factor in determining the performance of OCBT. Thus, in Section \ref{Sec_Performance}, we will use an eclectic filter that is selected with this trade-off in mind.

\section{Performance Evaluation} \label{Sec_Performance}

    \begin{figure}[b!]  
   \vspace{-0.2cm}
    \begin{center}
    {
	 \includegraphics[width=0.9\columnwidth]{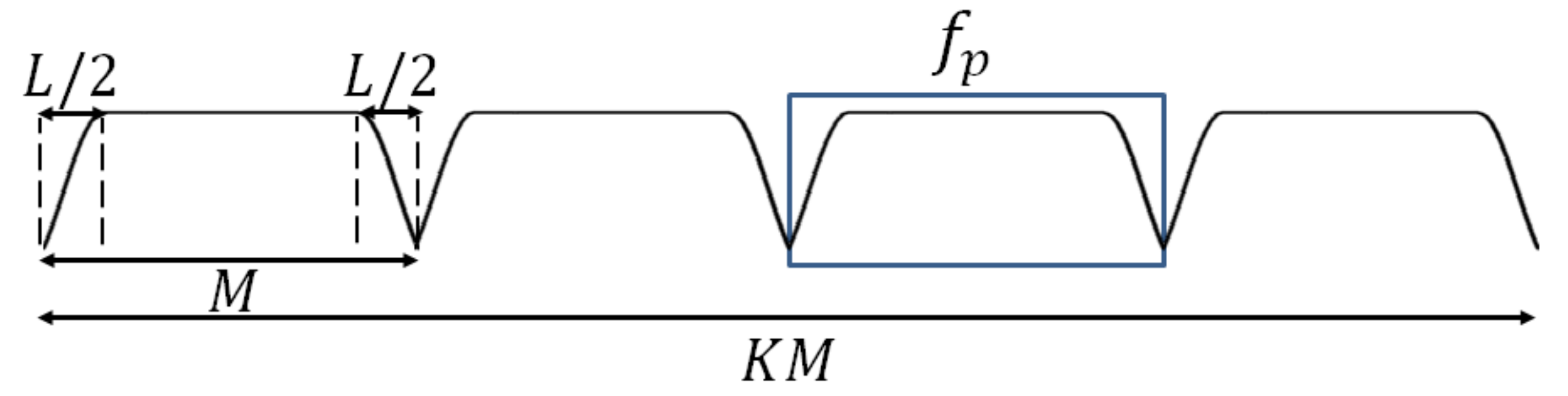}
   }
    \end{center}
    \vspace{-0.6cm}
    \caption{Window for OCBT using a raised cosine filter with $\beta=0.1$.}
   \label{Fig_window_OCBT}
   \end{figure}
   In this section, we compare the performances of OCBT with FBMC, CP-OFDM and W-OFDM in terms of the time efficiency, computational complexity,
   power spectral density (PSD) and bit error rate (BER). We set the number of sub-carriers $M$, the spreading and overlapping factor $K$ and $CP$ to 1024, 4 and $M/4=64$, respectively. In the PSD performance, we especially set $M$ to 64 to show the OOB radiation better.

   In OCBT and W-OFDM, we use one window consistently for each system to ensure the fair comparisons.
   We use a raised cosine filter at both ends of the window $f_p$ for OCBT that is sampled at $\frac{L}{2}$ samples/symbol with a roll-off factor ($\beta$) 0.1 and truncated to span 2 symbols. Considering the trade-off between the frequency confinement and the signal distortion of the window, we set the proportion of $L$ to $M$ as $324/1024$. The window $f(p{T_s})$ used here is depicted in Fig. \ref{Fig_window_OCBT}.

   Then, we use the raised cosine window for W-OFDM described in \cite{C4} that extends the CP to achieve the same resistance to ISI as CP-OFDM and uses an additional cyclic suffix (CS). Let us denote the CP of W-OFDM as $\text{CP}_\text W$. Both $\text{CP}_\text W$ and CS include the roll-off region which is overlapped with adjacent symbols with the same length $W$. Specifically, we set ${CP}_W = {\frac{3}{2}}{CP}$, ${CS} = \frac{1}{2}{CP}_W$ and $W = \frac{1}{3}{CP}_W$, where ${CP}$, ${CP}_W$ and $CS$ denote the length of the CP, $\text{CP}_\text W$ and CS, respectively.

   \subsection{Time Efficiency} \label{Subsec_Time_efficiency_performance}

    \begin{figure}[b!]  
    \vspace{-0.4cm}
    \begin{center}
    {
     \includegraphics[width=0.9\columnwidth]{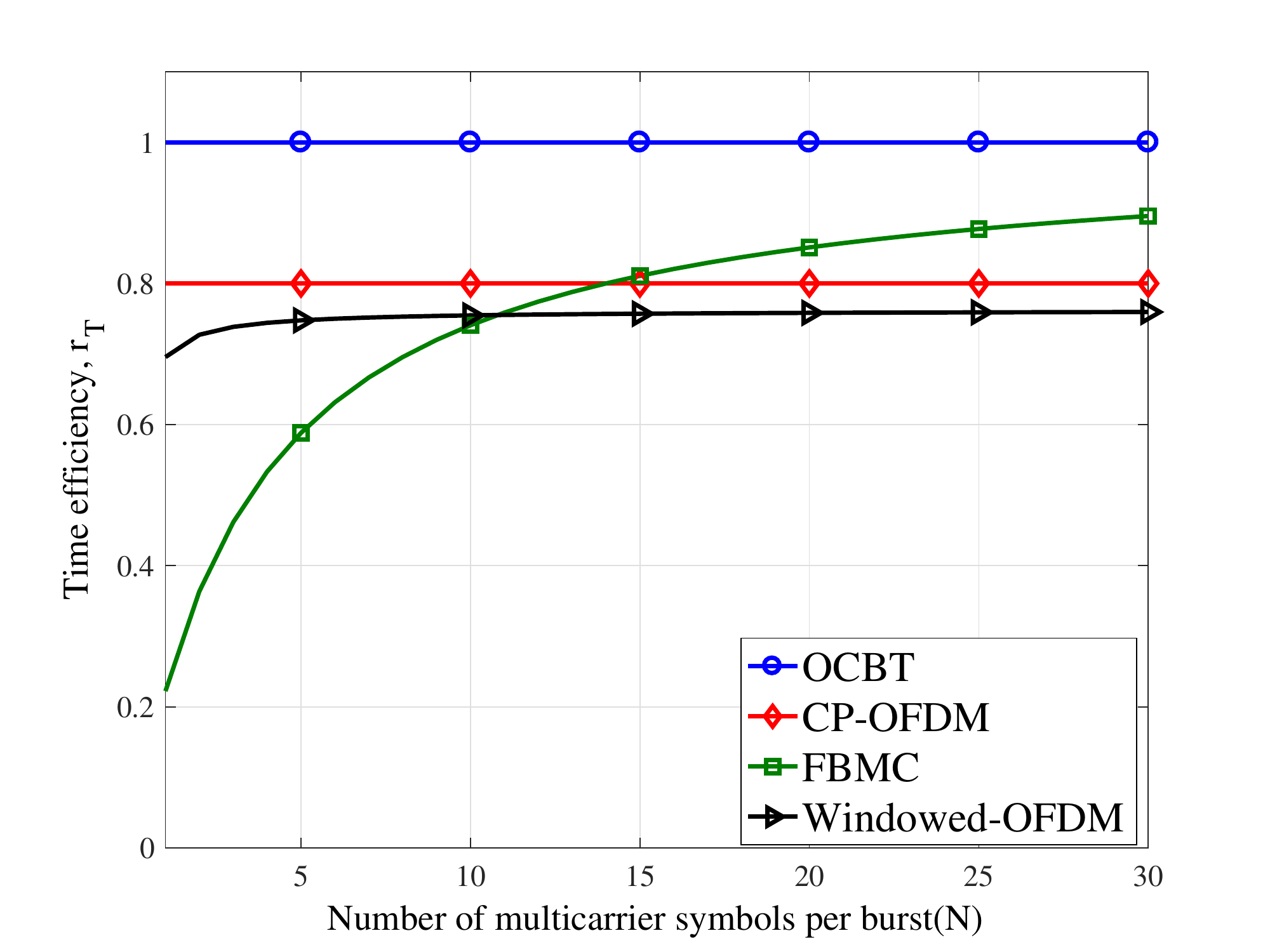}
    }
    \end{center}
    \vspace{-0.5cm}
    \caption{Time efficiency where $M=1024$, $K=4$, $CP=M/4$.}
    \label{Fig_time_efficiency}
    \end{figure}
    We evaluated the time efficiency of OCBT and compared it with the time efficiency of the other multi-carrier systems. The time efficiency \cite{C2}
    is defined as follows:
    \begin{equation}
    \label{Eq_time_eff}
    {r_T} = \frac{{{L_I}}}{{{L_I} + {L_T}}},
    \end{equation}
    where $L_I$ is the length of the transmitted information and $L_T$ is the length of the tail.
    $L_I$ is same as $MN$ for all of the systems because we can select $K$ for OCBT according to $N$, where $N$ multi-carrier symbols are transmitted with $M$ samples in each symbol.
    The length of the tails that make up the overhead are then described as
    \begin{equation}
    \label{Eq_L_T}
    \begin{array}{l}
    {L_{T,OCBT}} = 0,\\
    {L_{T,CP - OFDM}} = \left({CP}\right)N,\\
    {L_{T,FBMC}} = \frac{M}{2} + \left( {K - 1} \right)M,\\
    {L_{T,W - OFDM}} = \left({CP}_\text W + {CS}-{W} \right)N+{W}.
    \end{array}
    \end{equation}

    The time efficiency $r_T$ of the multi-carrier systems with $N$ symbols is shown in Fig. \ref{Fig_time_efficiency}. The time efficiency for OCBT is $1$, no matter how many symbols are transmitted, because OCBT does not need any tails, thanks to the time-spreading method with the orthogonal code. On the other hand, the time efficiency of the other systems cannot reach $1$ due to the overhead such as prefix, suffix and transition time. In particular, FBMC has the inferior time efficiency when transmitting the small number of symbols, because $L_{T,FBMC}$ in (\ref{Eq_L_T}) is fixed regardless of $N$.

    As shown in the Fig. \ref{Fig_time_efficiency}, OCBT has 25\% higher time efficiency than CP-OFDM and up to 350\% higher time efficiency than FBMC when one symbol is transmitted. In conclusion, OCBT is an outstanding candidate for improving time efficiency for future communication systems that require low overhead.

   \subsection{Power Spectral Density} \label{Subsec_PSD_performance}

   \begin{figure}[b!]
    \vspace{-0.4cm}
    \begin{center}
    {
	 \includegraphics[width=0.9\columnwidth]{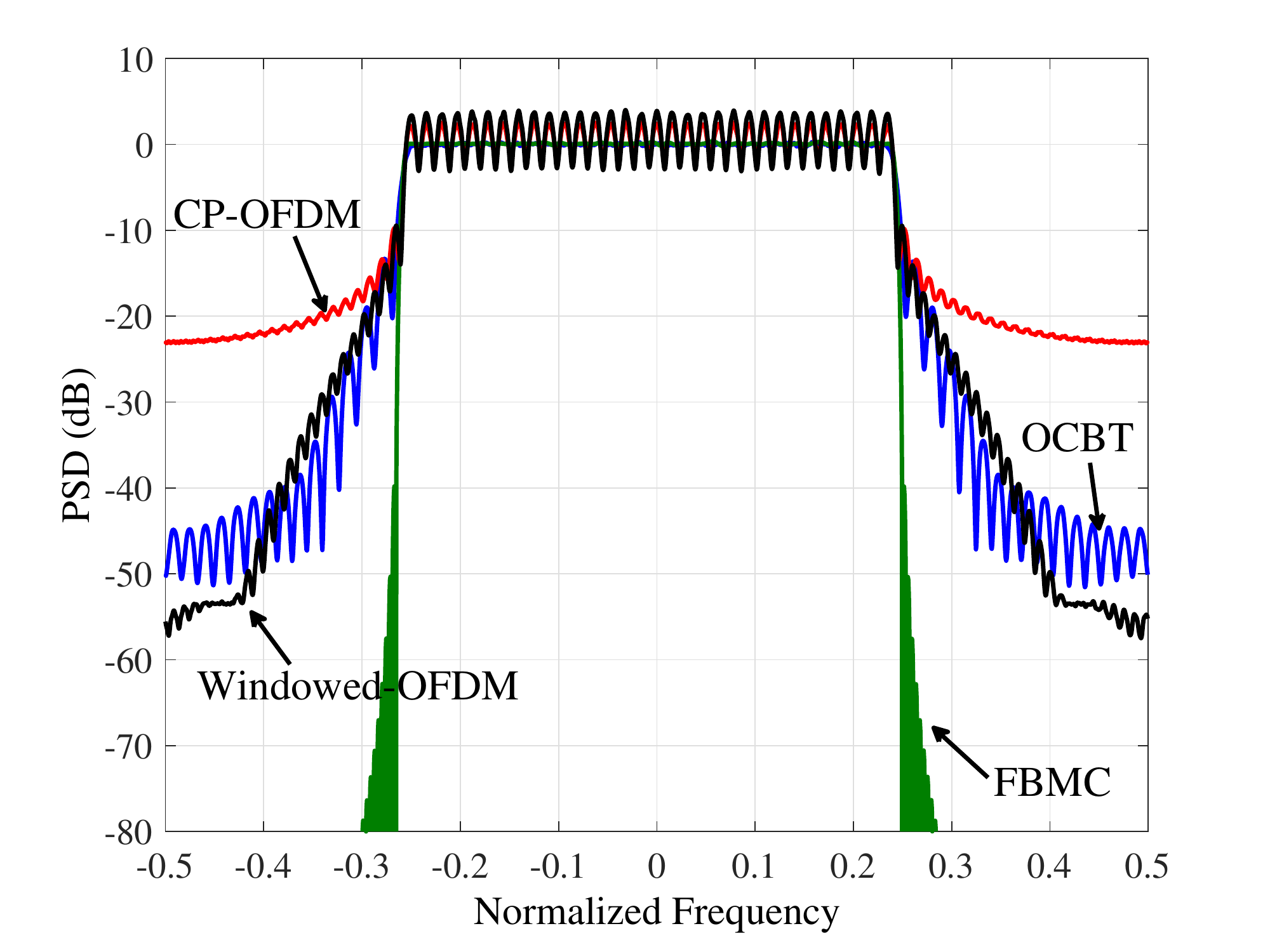} 
    }
    \end{center}
    \vspace{-0.5cm}
    \caption{PSD where $M=64$, $M/2$ active sub-carriers, $K=4$, $L=20$, $CP=M/4$.}
   \label{Fig_PSD}
   \end{figure}

    Fig. \ref{Fig_PSD} shows the PSD for the multi-carrier systems. We used $M/2$ active sub-carriers to see the radiation outside the active sub-carriers.
    The results show that FBMC outperforms the other schemes significantly because the prototype filter reduces the side-lobe level \cite{J5}.
    However, similarly to W-OFDM, OCBT also has the low OOB radiation with about a 30dB benefit compared to CP-OFDM, because windowing enhances the frequency localization effectively. Thus, OCBT reduces the required guard band so helping improve the spectral efficiency.

   \subsection{Computational Complexity} \label{Subsec_Computational Complexity_performance}

    We also investigate the complexity through the number of complex multiplications (CMs) required for the generation of one symbol of the multi-carrier systems as \cite{E4}.

    For the simple generation of one block of OCBT, $N$ symbols after $M$-point IFFT are first windowed by $f_p$, and then each windowed sub-block is added or subtracted according to the corresponding codes. Thus, OCBT needs additionally $NM$ CMs for one block only in windowing compared to OFDM, that is $M$ CMs for one symbol. The number of CMs of the other multi-carrier systems are summarized in Table \ref{Table_complexity}. Compared to FBMC and W-OFDM, OCBT has much lower complexity.
    \begin{table}[t!]  
    \begin{center}
    \caption{Computational complexity}
    \label{Table_complexity}
    \vspace{-0.1cm}
    \renewcommand{\arraystretch}{1.2}
    \begin{tabular}{l|c}
    \hline\hline
    Multi-carrier system & \begin{tabular}{@{}c@{}}Complexity expressions \\ where $M$=1024, $K$=4, CP=$M$/4\end{tabular}\\ \hline\hline
    OFDM & $\frac{M}{2}{\log_2}M$ = 5120\\ \hline
    FBMC    & $\frac{M}{2}{\log_2}M+(K+1)M$ = 10240\\  \hline
    W-OFDM & $\frac{M}{2}{\log_2}M+(M+{CP}_{w} + {CS})$ = 6720\\ \hline
    OCBT          & $\frac{M}{2}{\log_2}M+M$ = 6144\\ \hline\hline
    \end{tabular}
    \end{center}
    \vspace{-0.7cm}
    \end{table}

   \subsection{Bit Error Rate} \label{Subsec_BER_performance}

   \begin{figure}[b!]  
   \vspace{-0.4cm}
    \begin{center}
    {
	 \includegraphics[width=0.9\columnwidth]{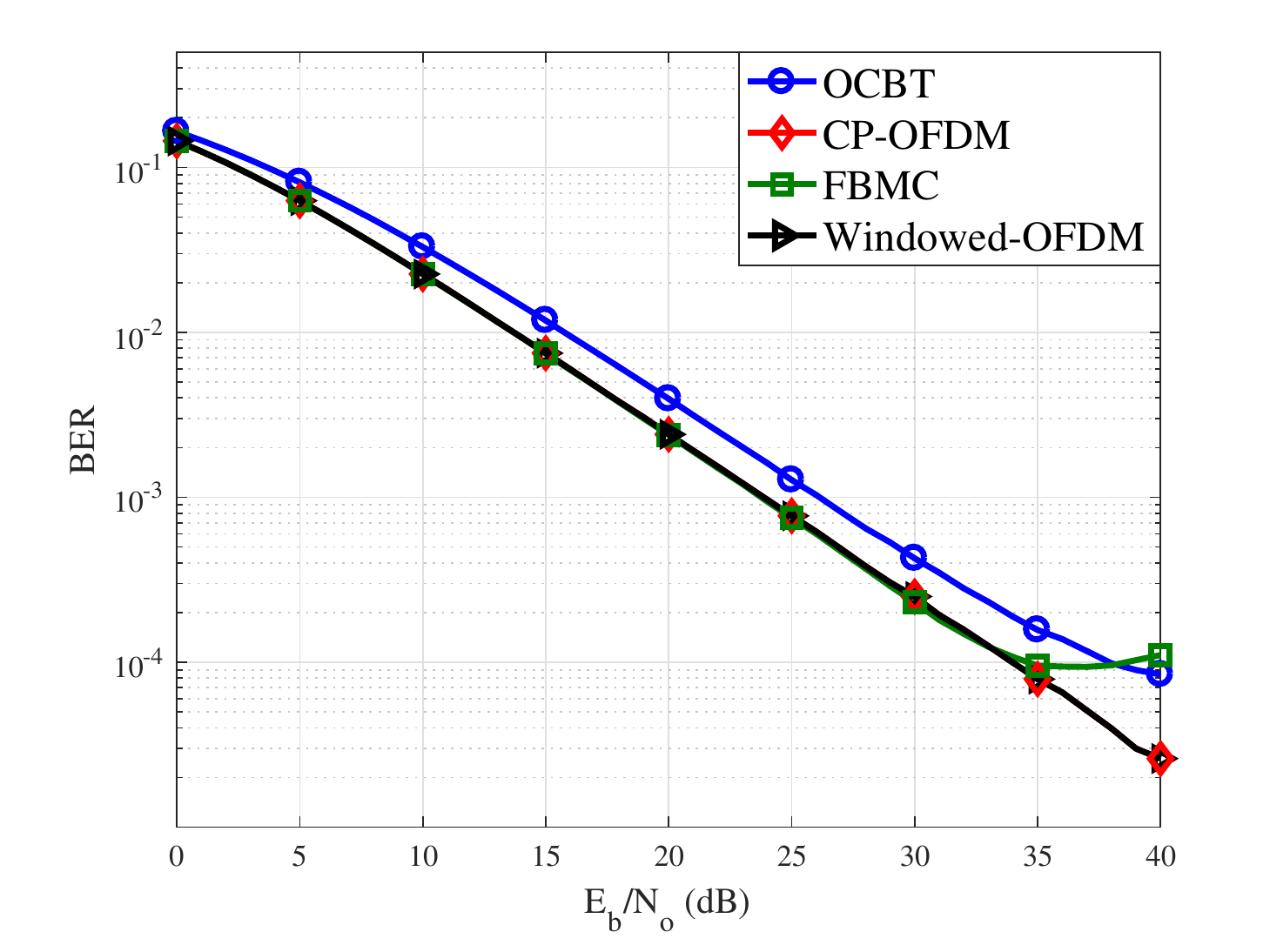}
   }
    \end{center}
    \vspace{-0.5cm}
    \caption{BER where $M=1024$, $K=4$, $L=324$, $CP=M/4$.}
   \label{Fig_BER}
   \end{figure}
   Let us consider the QPSK modulation except FBMC which uses OQAM and ITU-R vehicular-A channel model \cite{E3} with 2GHz center frequency, 30.72 MHz sampling frequency and 30km/h vehicular velocity. 
   The multi-carrier systems including OCBT use the MMSE equalizer \cite {J10} in the frequency domain.

   The BERs for OCBT, CP-OFDM, FBMC and W-OFDM are shown in Fig. \ref{Fig_BER}. As mentioned in (\ref{Eq_received_symbol_2}) in Section \ref{Sec_Performance_Analysis}, the received symbols of OCBT are distorted because of attenuation and ICI by the window. However, the BER degradation is a little because the IBI is reduced by the block structure with the time-spreading method even without the overhead.
   The OCBT has the BER performance degradation, however, unlike FBMC \cite{J11}, OCBT does not have an error floor in the high signal-to-noise ratio region as shown in Fig. \ref{Fig_BER}.

\section{Conclusions} \label{Sec_Con}

    In this paper, we proposed a new multi-carrier system, OCBT, and compared the performances of OCBT with CP-OFDM, FBMC, and W-OFDM. Comparisons with the other systems show that OCBT is robust to multi-path channels with higher time efficiency and lower computational complexity than FBMC because the block structure with the time-spreading method reduces the IBI. In addition, OCBT has about 30dB lower OOB radiation than CP-OFDM thanks to windowing. The OCBT can use QAM signal unlike FBMC, so it is easily applicable to the conventional QAM based MIMO techniques.
    Thus, OCBT, which can apply burst transmission more efficiently and apply QAM-based MIMO techniques more easily compared to FBMC, is the better choice for future communication systems.

    Topics for future works include investigating the optimal window in terms of both the low OOB radiation and little signal distortion for OCBT.

\bibliographystyle{reference}
\bibliography{IEEEabrv,reference}

\begin{thebibliography}{10}
\providecommand{\url}[1]{#1}
\csname url@samestyle\endcsname
\providecommand{\newblock}{\relax}
\providecommand{\bibinfo}[2]{#2}
\providecommand{\BIBentrySTDinterwordspacing}{\spaceskip=0pt\relax}
\providecommand{\BIBentryALTinterwordstretchfactor}{4}
\providecommand{\BIBentryALTinterwordspacing}{\spaceskip=\fontdimen2\font plus
\BIBentryALTinterwordstretchfactor\fontdimen3\font minus
  \fontdimen4\font\relax}
\providecommand{\BIBforeignlanguage}[2]{{%
\expandafter\ifx\csname l@#1\endcsname\relax
\typeout{** WARNING: IEEEtran.bst: No hyphenation pattern has been}%
\typeout{** loaded for the language `#1'. Using the pattern for}%
\typeout{** the default language instead.}%
\else
\language=\csname l@#1\endcsname
\fi
#2}}
\providecommand{\BIBdecl}{\relax}
\BIBdecl

\bibitem{E5}
\textrm{Dahlman, Erik, Stefan Parkvall, and Johan Skold.}, ``\textrm{4G:
  LTE/LTE-advanced for mobile broadband},'' \emph{Academic press}, 2013.

\bibitem{C3}
M.~Batariere, K.~Baum, and T.~P. Krauss., ``\textrm{Cyclic prefix length
  analysis for 4G OFDM systems},'' in \emph{IEEE 60th Vehicular Technology
  Conference (VTC 2004-Fall)}, vol.~1, Sep. 2004, pp. 543--547.

\bibitem{E2}
\textrm{M. Bellanger et al.}, ``\textrm{FBMC Physical Layer: A Primer},'' Jun.
  2010.

\bibitem{J4}
P.~Siohan, C.~Siclet, and N.~Lacaille., ``\textrm{Analysis and design of
  OFDM/OQAM systems based on filterbank theory},'' \emph{IEEE Transactions on
  Signal Processing}, vol.~50, no.~5, pp. 1170--1183, May. 2002.

\bibitem{J9}
H.~Nam, M.~Choi, S.~Han, C.~Kim, S.~Choi, and D.~Hong., ``\textrm{A New
  Filter-Bank Multicarrier System With Two Prototype Filters for QAM Symbols
  Transmission and Reception},'' \emph{IEEE Transactions on Wireless
  Communications}, vol.~15, no.~9, pp. 5998--6009, Sep. 2016.

\bibitem{C5}
H.~S. Sourck, Y.~Wu, J.~W.~M. Bergmans, S.~Sadri, and B.~Farhang-Boroujeny.,
  ``\textrm{Effect of carrier frequency offset on offset qam multicarrier
  filter bank systems over frequency-selective channels},'' in \emph{2010 IEEE
  Wireless Communication and Networking Conference}, Apr. 2010, pp. 1--6.

\bibitem{C4}
T.~Weiss, J.~Hillenbrand, A.~Krohn, and F.~K. Jondral., ``\textrm{Mutual
  interference in OFDM-based spectrum pooling systems},'' in \emph{IEEE 59th
  Vehicular Technology Conference (VTC 2004-Spring)}, vol.~4, May 2004, pp.
  1873--1877.

\bibitem{C2}
F.~Schaich, T.~Wild, and Y.~Chen., ``\textrm{Waveform Contenders for 5G -
  Suitability for Short Packet and Low Latency Transmissions},'' in \emph{IEEE
  79th Vehicular Technology Conference (VTC 2014-Spring)}, May 2014, pp. 1--5.

\bibitem{J7}
Z.~Wang and G.~B. Giannakis., ``\textrm{Wireless multicarrier
  communications},'' \emph{IEEE Signal Processing Magazine}, vol.~17, no.~3,
  pp. 29--48, May. 2000.

\bibitem{J10}
P.~Tan and N.~C. Beaulieu., ``\textrm{A Comparison of DCT-Based OFDM and
  DFT-Based OFDM in Frequency Offset and Fading Channels},'' \emph{IEEE
  Transactions on Communications}, vol.~54, no.~11, pp. 2113--2125, Nov. 2006.

\bibitem{J8}
A.~Sahin and H.~Arslan., ``\textrm{Edge Windowing for OFDM Based Systems},''
  \emph{IEEE Communications Letters}, vol.~15, no.~11, pp. 1208--1211, Nov.
  2011.

\bibitem{J5}
B.~Farhang-Boroujeny., ``\textrm{OFDM Versus Filter Bank Multicarrier},''
  \emph{IEEE Signal Processing Magazine}, vol.~28, no.~3, pp. 92--112, May.
  2011.

\bibitem{E4}
\textrm{Luo, Fa-Long, and Charlie Zhang.}, ``\textrm{Signal Processing for 5G:
  Algorithms and Implementations},'' \emph{John Wiley \& Sons}, 2016.

\bibitem{E3}
\textrm{Recommendation ITU-R M.1225}, ``\textrm{Guidelines for evaluation of
  radio transmission technologies for IMT-2000},'' \emph{Int. Telecommun.
  Union}, 1997.

\bibitem{J11}
Z.~Koll{\'a}r, J.~Bit{\'o}, L.~Varga, and P.~Horv{\'a}th, ``\textrm{Novel
  multicarrier modulation for cognitive radio systems},'' \emph{IEICE TCSR,
  Fukuoka, Japan, IEICE Tech. Rep.}, vol. 112, p.~67, 2012.

\end{thebibliography}
\end{document}